\documentclass[
    ,final            
  ]
  {aipproc}

\usepackage{amsmath,amssymb,graphicx,slashbox}

\layoutstyle{8x11double}

\def\mydate{September 12, 2008}
\def\ignore#1{{}}

\newcommand{\beeq}{\begin{equation}}
\newcommand{\eneq}{\end{equation}}
\newcommand{\beqn}{\begin{eqnarray}}
\newcommand{\eeqn}{\end{eqnarray}}

\def\dd{\partial}
\def\la{\raise.16ex\hbox{$\langle$}\lower.16ex\hbox{}  }
\def\ra{\, \raise.16ex\hbox{$\rangle$}\lower.16ex\hbox{} }
\def\go{\rightarrow}

\def\onehalf{ \hbox{${1\over 2}$} }

\def\Tr{{\rm Tr \,}}

\def\eff{{\rm eff}}

\def\EM{{\rm EM}}
\def\diag{{\rm diag ~}}

\def\KK{{\rm KK}}

\def\Psibar{\overline{\Psi}}

\begin{document}

\title{Electroweak Gauge-Higgs Unification Scenario}

\classification{11.10.Kk, 11.15.Ex, 12.60.-i, 12.60.Cn
               \hfill {\bf Preprint number:} OU-HET 611/2008}
\keywords      {gauge-Higgs unification, Hosotani mechanism \hfill \mydate}

\author{Yutaka Hosotani}{
  address={Department of Physics, Osaka University, Toyonaka, Osaka 560-0043, Japan}
}

\begin{abstract}
In the gauge-Higgs unification scenario 4D Higgs fields are unified
with gauge fields in higher dimensions.   The electroweak model is
constructed in the Randall-Sundrum warped space.  The electroweak
symmetry is dynamically broken by the Hosotani  mechanism due to
the top quark contribution.  The Higgs mass is predicted to be around
50 GeV with the vanishing $ZZH$ and $WWH$ couplings so that
the LEP2 bound for the Higgs mass is evaded.
\end{abstract}

\maketitle


\subsection{Introduction}

We are in the hunt for the Higgs particle. 
It  is expected to be discovered at LHC in the near future.
In the standard model of the electroweak interactions
the Higgs particle is necessary to induce the electroweak symmetry breaking, 
but  it is not obvious if the Higgs particle appears as described
in the standard model.   
\ignore{Its mass and  couplings to other particles
may deviate from those in the standard model.  }

The Higgs sector in the standard model is not completely satisfactory.
It lacks underlying principles.  Is there  a principle governing the Higgs field?
What is the origin of the Higgs particle?  
After all, what is the mechanism of the electroweak gauge symmetry 
breaking?

The gauge-Higgs unification scenario tries to answer to these 
questions.\cite{Fairlie1}-\cite{YH2}
It identifies the Higgs field as a part of gauge fields in higher dimensional 
theory.  The 4D Higgs field is unified with gauge fields, the Higgs interactions
being controlled by the gauge principle, once the back-ground spacetime 
is specified.  In this scenario the electroweak symmetry is dynamically 
broken, the Higgs mass being predicted at a finite value.  
The Higgs couplings to the $W$ and $Z$ bosons and fermions deviate
from those in the standard model, which can be checked experimentally 
in the near future.

\subsection{EW Gauge-Higgs unification}

We consider a gauge theory defined in higher dimensions with 
not-simply-connected extra-dimensional space.\cite{YH1, YH2}  
There appear  Aharonov-Bohm (AB) phases along the extra dimension, 
which, though giving vanishing field strengths, become  physical degrees
of freedom.  4D Higgs fields are nothing but  4D fluctuation modes of
such AB phases.  
In terms of gauge potentials $A_M(x,y) = (A_\mu, A_y)$, 
AB phases are generated by
zero modes of the extra-dimensional components $A_y (x,y)$.
In non-Abelian gauge theory those AB phases can 
develop non-trivial vacuum expectation values at the quantum level,
inducing dynamical gauge symmetry breaking.  

This scenario has many attractive features.  
The Higgs particle is massless at the tree level, as AB phases $\theta_H$ 
give vanishing field strengths.  
The effective potential for the AB phases $V_\eff (\theta_H)$, however, 
becomes nontrivial at the quantum level.  The Higgs mass, which is 
proportional to the curvature of $V_\eff$ at its global minimum, is
generated.  Thanks to the gauge invariance the mass is predicted at a
finite value, irrespective of an ultra-violet cutoff introduced,
which can be used for solving the gauge hierarchy problem.\cite{Lim2}
Further the Higgs couplings are determined by the gauge principle.

There are several key ingredients in applying the scenario to the electroweak
interactions.

\noindent
{\bf 1. Larger gauge group $G$}

In the  EW symmetry breaking $SU(2)_L \times U(1)_Y \go U(1)_\EM$
the Higgs field is an $SU(2)_L$ doublet.  In the gauge-Higgs unification 
the Higgs field is a part of gauge fields which are in the adjoint representation
of the gauge group $G$.  This implies that one needs to start with a larger
gauge group $G$ which contains  $SU(2)_L \times U(1)_Y$ as a subgroup.
Examples are $SU(3)$, $SU(3) \times U(1) \times U(1)$, and 
$SO(5) \times U(1)$.

\noindent
{\bf 2. Orbifolds}

As an extra-dimensional space we take an orbifold.\cite{Pomarol1}
The simplest example is
$S^1/Z_2$ in which the points $y$, $y+2\pi R$, and $-y$ are identified.
Physics must be the same at those points, but gauge potentials need not. 
Gauge potentials obey, around two fixed points $y_0=0$ and
$y_1= \pi R$, 
\beeq
\begin{pmatrix}  A_\mu  \\  A_y  \end{pmatrix} ( x, y_j - y) 
= P_j  \begin{pmatrix}  A_\mu  \\  -A_y  \end{pmatrix} ( x, y_j +y) P_j^\dagger ~,
\label{BC1}
\eneq
where $P_j = P_j^{-1} \in G$.  It follows that 
$A_M(x, y + 2\pi R) = U A_M(x, y) U^\dagger$ where $U = P_1 P_0$.  
The Lagrangian density remains invariant under the parity transformations.
The set $\{ P_0, P_1 \}$ defines the orbifold boundary conditions (BC).

\noindent
{\bf 3. Four-dimensional  Higgs}

4D Higgs fields reside in the $A_y$ components which are even under
$P_0$ and $P_1$.  
Take $G= SO(5)$ and $P_0= P_1= {\rm diag}~(-1, -1, -1, -1, 1)$.
With this orbifold BC $SO(5)$ breaks down to $SO(4)$.   
$A_\mu$'s have zero modes (4D gauge fields)  in the diagonal 
$SO(4) \simeq SU(2)_L \times SU(2)_R$.
$A_y$, on the other hand,  has zero modes in the off-diagonal parts 
with $*$ in
\beeq
A_y \sim 
\begin{pmatrix} &&&&* \cr &&&&* \cr &&&&* \cr &&&&* \cr * &* &* &* 
\end{pmatrix} ~. 
\label{Ay1}
\eneq
The zero mode multiplet is  an $SO(4)$ vector, or a $({\bf 2}, {\bf 2})$ 
representation of
$SU(2)_L \times SU(2)_R$.  It can be identified with the EW Higgs field.
 
The 4D Higgs field naturally appears from the orbifold BC.  The AB phase,
or the Wilson line phase, is given by
\beeq
e^{i \Theta_H/2} 
=P \exp \Big\{ i g \int_{y_0}^{y_1} dy  ~  A_y \Big\} ~.
\label{AB1}
\eneq

\noindent
{\bf 4. Chiral fermions}

Another virtue of the orbifold structure is that it naturally gives rise to chiral
fermions.  Take a vector fermion multiplet $\Psi$ in the  $SO(5)$ model 
above.   The orbifold BC for $\Psi$ is given by
\beeq
\Psi(x, y_j -y) = \pm P_j  \gamma^5 \Psi(x, y_j + y) ~.
\label{BC2}
\eneq
The factor $\gamma^5$ is necessary to assure the invariance of
$\overline{\Psi}  i ( \gamma^\mu D_\mu +  \gamma^5 D_5 ) \Psi$. 
With $+$ sign in (\ref{BC2}), the first four components of $\Psi$ have
zero modes only for $\gamma^5=-1$ (left-handed components), 
whereas the fifth component has a zero mode only for $\gamma^5 = 1$ 
(a right-handed component).  All the massive Kaluza-Klein excited
states appear vector-like, but the lowest, light modes appear chiral.

\noindent
{\bf 5. Flat v.s. warped}

With all virtues of orbifolds, many models of electroweak interactions
have been constructed.\cite{Antoniadis1}-\cite{Csaki2}
The value of the Wilson line phase $\theta_H$ 
is determined once the matter content is specified.  With simple, minimal 
matter content the effective potential $V_\eff (\theta_H)$ is minimized,
typically, either at $\theta_H =0$ or at $\theta_H = O(1)$.  In the former 
case the EW symmetry is unbroken, whereas in the latter case the symmetry 
generally breaks.

In models in flat space, say, on $M^4 \times (S^1/Z_2)$, the Kaluza-Klein
scale is given by $m_\KK = 1/R$ where $R$ is the radius of $S^1$.  
With  $\theta_H = O(1)$ the $W$ boson acquires a mass 
$m_W \sim (\theta_H/2\pi) m_\KK$, which leads to a too low $m_\KK$.
Secondly, the Higgs mass is generated at the one loop level so that
$m_H \sim \sqrt{\alpha_W} \, (2\pi/\theta_H) m_W$ where 
$\alpha_W = g_W^2/4\pi$.  This leads to a too small $m_H \sim 10 \,$GeV.
Thirdly, the $WWZ$ coupling deviates from the value in the standard model 
as the wave functions of $W$ and $Z$ in the fifth dimension $y$ acquire
significant dependence on $y$, which contradicts with the LEP2 experiment.

There are two approaches to solve these problems.  One way is to stay 
in flat space and tune the matter content such that $V_\eff (\theta_H)$
is minimized at a small value for $\theta_H$.  For instance, one can 
introduce many matter multiplets,  or even supersymmetry,  to have cancellation 
among dominant parts of the contributions to  $V_\eff$.
Or, one can incorporate quarks in several representations of the gauge group
to have small $\theta_H$.

An alternative way is to consider models in the curved space, particularly
in the Randall-Sundrum (RS) warped space.\cite{HM}  
It is remarkable that all the 
problems mentioned above are naturally solved in the RS space.

\subsection{Models in the warped space}

The metric in the Randall-Sundrum (RS) warped spacetime is given by
\beeq
 ds^2 
 = e^{-2\sigma (y)} \, \eta_{\mu\nu} dx^\mu dx^\nu
 +dy^2 ~~, 
 \label{metric1}
\eneq
where $\eta_{\mu\nu} = \diag(-1,1,1,1)$, $\sigma (y)=\sigma (y+2L)$, 
and $\sigma (y) \equiv k |y|$ for $|y| \leq  L$.   The extra-dimensional space has 
the topology of $S^1/Z_2$.  The fundamental region is given by $0 \le y \le L$.
The bulk region $0 <y < L$, which is a sliced AdS spacetime with a 
negative cosmological constant $\Lambda = - 6 k^2$, 
is sandwiched by the Planck brane at $y=0$ and the TeV brane at $y=L$. 
A large warp factor $z_L = e^{kL} \gg 1$ bridges the Planck scale  and
the weak scale.
The KK mass scale for fields defined in the bulk is 
\beeq
m_\KK = \frac{\pi k}{e^{kL} -1} \sim \pi k e^{-kL} ~.
\eneq

\noindent
{\bf 1. Gauge group}

At the moment the  most promising model is a model based on 
$SO(5) \times U(1)_X$ gauge symmetry.\cite{Agashe2, SH1, HS2}  
The orbifold boundary condition
$\{ P_0, P_1 \}$ is given by $P_0= P_1= {\rm diag}~(-1, -1, -1, -1, 1)$
in the $SO(5)$ part, which reduce  the residual symmetry to
$SO(4) \times U(1)_X \simeq SU(2)_L \times SU(2)_R \times U(1)_X$.  
On the Planck brane the symmetry $SU(2)_R \times U(1)_X$ is further 
spontaneously broken by additional brane dynamics to $U(1)_Y$. 
The residual symmetry is $SU(2)_L \times U(1)_Y$.  

\noindent
{\bf 2. Higgs field}

There appear zero modes in $A_y$ as shown in (\ref{Ay1}),  
which become the 4D Higgs fields $\Phi_H(x)$  in the EW theory.  
The relation in $0 \le y \le L$ is given by
\beqn
&&\hskip -1cm
A_y^{a5} (x, y) = \phi^a (x)  \sqrt{\frac{2k}{z_L^2 -1}} \, e^{2ky} 
+ \cdots ~,  \cr
&&\hskip -1cm
\Phi_H(x) = \frac{1}{\sqrt{2}}
\begin{pmatrix} \phi^2 + i \phi^1 \cr \phi^4 - i \phi^3 \end{pmatrix} ~.
\label{Ay2}
\eeqn
We note that the Higgs field is localized near the TeV brane in the warped space. 
When $\la \phi^a \ra = v \delta^{a4}$, the Wilson line phase is given by
\beeq
\theta_H = \frac{g_A v}{2} \sqrt{\frac{z_L^2 -1}{k}} 
\label{Wilson2}
\eneq
where $g_A$ is the gauge coupling of $SO(5)$ related to the 4D
$SU(2)_L$ weak coupling by $g_W = g_A/ \sqrt{L}$.  

\noindent
{\bf 3. Photon, $W$ and $Z$}

$U(1)_\EM$ remains as an exact symmetry.  The photon wave function is
constant in the $y$ coordinate, being completely flat and independent of 
$\theta_H$.  With $\theta_H \not= 0$ the EW symmetry breaks and 
$W$ and $Z$ become massive.  For large $z_L$ 
\beeq
m_W \sim  \sqrt{\frac{k}{L}} \, e^{-kL}  \, | \sin\theta_H | 
\sim  \frac{m_\KK}{\pi \sqrt{kL} } \, | \sin\theta_H |  ~~.
\label{Wmass}
\eneq
With $|\sin \theta_H | = O(1)$ one finds that for 
$z_L \sim 10^{15} (10^{17})$, 
$k\sim 5 \times 10^{17} ~ (5 \times 10^{19}) \, $GeV and
$m_\KK \sim 1.5 ~ (1.6) \, $TeV.  
The $Z$ boson mass is given by
\beeq
m_Z \sim \frac{m_W}{\cos \theta_W} ~~,~~
\sin \theta_W = \frac{g_B}{\sqrt{g_A^2 + 2 g_B^2}} 
\label{Zmass}
\eneq
where $g_B$ is the gauge coupling of $U(1)_X$.

As the EW symmetry breaks down, the wave functions of $W$ and $Z$ 
acquire $y$-dependence.  However, as $\la A_y \ra$ or the Higgs field 
is localized near the TeV brane, the $W$ and $Z$ wave functions remain
almost flat in the bulk, non-trivial $y$-dependence appearing only near 
the TeV brane.  

\noindent
{\bf 4. Gauge self-couplings}

In the standard model the gauge coupling is universal. 
Gauge couplings of quarks, leptons  and the Higgs field
as well as $WWZ$, $WWWW$, $WWZZ$  self-couplings
are all specified with the two gauge coupling constants of
$SU(2)_L$ and $U(1)_Y$.  In higher dimensional theory 
this is not the case anymore.  The reason is that four-dimensional
gauge couplings are determined by overlap integrals of 
wave functions of associated fields in the extra dimension.
The only exactly-universal coupling is the electromagnetic
coupling associated with the exact symmetry of $U(1)_\EM$.

This poses us a challenging test for higher dimensional theory.
The $WWZ$ coupling has been already measured  indirectly
at LEP2.  The coupling agrees with that in the standard model 
within a few percents.

In the gauge-Higgs unification scenario the gauge self-couplings
and gauge-Higgs couplings in four dimensions are determined 
from the $\Tr \, F_{MN} F^{MN}$ term.  In particular the part
$\Tr \, F_{\mu\nu} F^{\mu\nu}$ ($\mu,\nu=0, \cdots, 3$) gives
the $WWZ$, $WWWW$, $WWZZ$  couplings.  At $\theta_H=0$ 
these couplings reduce to those in the standard model.

In the Randall-Sundrum spacetime, as mentioned above, the wave
functions of $W$ and $Z$ remain almost $y$-independent even at
$\theta_H \not= 0$, though the weight in the group components
have significant $\theta_H$ dependence.  Thanks to this property 
the $WWZ$, $WWWW$, $WWZZ$  couplings remain almost universal.
For instance, the deviation of the  $WWZ$ coupling from that in 
the standard model is about $4 \times 10^{-5}$ or $2 \times 10^{-4}$
at $\theta_H = \pi/4$ or $\pi/2$, respectively.  

We remark that in the flat spacetime limit $k \go 0$ the deviation becomes
substantial as the wave functions acquire significant $y$-dependence.
The deviation becomes as large as 7\% at $\theta_H = \pi/2$, which already
contradicts with the LEP2 data.

\noindent
{\bf 5. Gauge-Higgs couplings}

The 4D Higgs field ($H$) is contained in $A_y$.  Overlap integrals of
the $\Tr \, F_{\mu y} F^{\mu y}$ term give the $WWH$, $ZZH$,
$WWHH$ and $ZZHH$ couplings.   The Higgs wave function is localized
near the TeV brane so that these couplings sensitively depend on the
behavior of the $W$ and $Z$ wave functions in the vicinity of the TeV
brane, and therefore on $\theta_H$.  

The result is robust. These couplings are given by
\beqn
\lambda_{WWH} &\simeq& g_W m_W \cos \theta_H  ~, \cr
\noalign{\kern 6pt}
\lambda_{ZZH} &\simeq& \frac{g_W m_Z}{\cos \theta_W}  \cos \theta_H ~,  \cr
\noalign{\kern 3pt}
\lambda_{WWHH}^{\rm bare} &\simeq&
 g_W^2  \Big(1 -  \frac{2}{3} \sin^2 \theta_H \Big) ~, \cr
\noalign{\kern 3pt}
\lambda_{ZZHH}^{\rm bare} &\simeq&
\frac{g_W^2}{\cos^2 \theta_W} \Big(1 -  \frac{2}{3} \sin^2 \theta_H \Big) ~.
\label{Higgs1}
\eeqn
The factor $\cos \theta_H$ in $\lambda_{WWH}$ and $\lambda_{ZZH}$
gives significant suppression compared with the couplings in the standard model.
The suppression can be measured at LHC, once the Higgs particle is found.

All of the $W$, $Z$ and $H$ bosons have their Kaluza-Klein (KK) towers. 
The KK excited states also have nontrivial gauge couplings.   It is shown that
the $WWH^{(n)}$ and $ZZH^{(n)}$ couplings identically vanish. 
The $WW^{(n)}H$ and $ZZ^{(n)}H$ couplings are substantial, however.
In the $WWHH$ coupling, for instance, $W^{(n)}$ can appear as an 
intermediate state ($WH \go W^{(n)} \go WH$) .  
It gives an important contribution to the low energy
effective $\lambda_{WWHH}^{\rm eff}$ coupling as described below.

\noindent
{\bf 6. Effective Lagrangian}

It is convenient to write down the effective Lagrangian at low energies
in terms of low-energy fields,  integrating over heavy fields (KK excited states).
This can be consistently carried out in the RS warped space, by utilizing
the holographic property as shown by Panico and Wurzer\cite{Panico2} 
and by Sakamura.\cite{Sakamura1}

The effective Lagrangian describing couplings of $H$
to  $W$, $Z$ is given by
\beqn
&&\hskip -1.cm
{\cal L}_\eff \sim
- \frac{g_W^2 f_H^2}{2}  
\sin^2 \Big( \theta_H + \frac{H}{\sqrt{2} f_H} \Big)\cr
&&\hskip 2.cm 
\times 
\Big\{ W_\mu^\dagger W^\mu 
+ \frac{Z_\mu Z^\mu}{2 \cos^2 \theta_W} \Big\} , \cr
\noalign{\kern 5pt}
&&\hskip -1.cm
f_H = \frac{1}{g_W} \sqrt{ \frac{2k}{L} } ~  e^{-kL}
= \sqrt{\frac{2}{kL}}  ~ \frac{m_\KK}{\pi g_W} ~.
\label{effLag1}
\eeqn
Notice that ${\cal L}_\eff$ is periodic in $\theta_H + (H/\sqrt{2} f_H)$.

Expanded in a Taylor series in $H$, ${\cal L}_\eff$ yields the mass terms 
for $W$ and $Z$ and the Higgs couplings to $W$ and $Z$.
$m_W$, $m_Z$, and $\lambda_{WWH}$ and  $\lambda_{ZZH}$ in 
(\ref{Higgs1}) are reproduced.  The couplings $\lambda_{WWHH}$ and  
$\lambda_{ZZHH}$, however, deviate from those in (\ref{Higgs1}), 
as they incorporate contributions from intermediate $W^{(n)}$ and 
$Z^{(n)}$.  They are found to be
\beqn
\lambda_{WWHH}^{\rm eff} &\simeq&
 g_W^2  \cos 2 \theta_H ~, \cr
\noalign{\kern 5pt}
\lambda_{ZZHH}^{\rm eff} &\simeq&
\frac{g_W^2}{\cos^2 \theta_W} \cos 2 \theta_H  ~.
\label{Higgs2}
\eeqn
The suppression factor, compared with the values in the standard
model, is given by $\cos 2 \theta_H$.

\noindent
{\bf 7. Tree unitarity in $W_LW_L$ scattering}

In the standard model the Higgs field plays an important role
to restore the unitarity in the elastic scattering of the longitudinal
components of $W$ and $Z$.  
In the gauge-Higgs unification scenario, however, 
the $\lambda_{WWH}$ and $\lambda_{ZZH}$ couplings  are
suppressed by a factor $\cos \theta_H$.  
If this is the case,  one might wonder if the unitarity is destroyed as
the contribution from the Higgs field exchange is diminished.  

This problem is analyzed in ref.\ \cite{Falkowski2}.  It is shown that
the decrease in the Higgs contribution is compensated by contributions
from $W^{(n)}$ or $Z^{(n)}$ exchange so that the tree unitarity is
maintained.

\subsection{Quarks and leptons}

Let us adopt the viewpoint that the observed quarks and leptons live
in the bulk five-dimensional spacetime.  A fermion in the bulk is described by
\beqn
&&\hskip -1.cm 
\Psibar \, \bigg\{ \Gamma^A {e_A}^M \Big(
\dd_M - \frac{1}{8} \omega_{MBC} [\Gamma^B, \Gamma^C]  \cr
&&\hskip .5cm 
 - ig_A A_M -i q \frac{g_B}{2}  B_M \Big)
  -  c \sigma'(y) \bigg\} \, \Psi 
\label{Lag4}
\eeqn
where  ${e_A}^M$ and $\omega_{MBC}$ are tetrads and spin connections.
$c$ is a dimensionless bulk mass parameter, which controls wave functions of
quarks and leptons.\cite{GP}
Implementing quarks and leptons in the $SO(5) \times U(1)_X$ model
is not  trivial as one has, in general, additional light exotic fermions.  
They have to be made heavy by some means.

\noindent
{\bf 1. Medina-Shar-Wagner (MSW) model}

In the quark sector three $SO(5)$ multiplets per generation are
introduced.\cite{Wagner1}  For the third generation one has
\beeq
{\bf 5}_1 = \begin{pmatrix} t_L \cr b_L \cr \vdots \end{pmatrix}
~,~
{\bf 5}_2 = \begin{pmatrix} t_R' \cr  \vdots \end{pmatrix}
~,~
{\bf 10} = \begin{pmatrix} b_R' \cr  \vdots \end{pmatrix}
\eneq
where light modes are denoted in the parentheses.  With the orbifold
boundary condition $\{ P_0, P_1 \}$ alone there appear 20 light modes.
16 unwanted modes are made heavy by assigning flipped boundary conditions.
Further brane masses are introduced at the TeV brane.  

This model is consistent with the
electroweak precision measurements.  It is shown that the model leads to
dynamical electroweak symmetry breaking in a wide range in the parameter 
space.

\noindent
{\bf 2. HOOS model}

A model with simpler matter content has been proposed.\cite{HOOS} 
For the third generation one has two {\bf 5} multiplets in the bulk
with  bulk mass parameters $c_1$ and $c_2$
\beqn
&&\hskip -1cm
\begin{pmatrix} T \cr B \cr t \cr b \cr t'  \end{pmatrix}
\Rightarrow 
Q_{1L}= \begin{pmatrix} T_L \cr B_L \end{pmatrix} ~,~
q_{L}= \begin{pmatrix} t_L \cr b_L \end{pmatrix} ~,~  t_R' ~, 
\cr
&&\hskip -1cm
\begin{pmatrix} U \cr D \cr X \cr Y \cr b'  \end{pmatrix}
\Rightarrow 
Q_{2L}= \begin{pmatrix} U_L \cr D_L \end{pmatrix} ~,~
Q_{3L}= \begin{pmatrix} X_L \cr Y_L \end{pmatrix} ~,~ b_R' ~, 
\label{bulkF}
\eeqn
and fermions living on the Planck brane which come in three right-handed 
multiplets  belonging to $({\bf 2}, {\bf 1})$ representation of 
$SU(2)_L \times SU(2)_R$
\beeq
\hat \chi_{1R} = \begin{pmatrix} \hat T_R \cr \hat B_R \end{pmatrix} ~,~
\hat \chi_{2R} = \begin{pmatrix} \hat U_R \cr \hat D_R \end{pmatrix} ~,~
\hat \chi_{3R} = \begin{pmatrix} \hat X_R \cr \hat Y_R \end{pmatrix} ~.
\label{braneF}
\eneq
The bulk fermions obey the normal orbifold boundary conditions.  On the right
side of each {\bf 5} multiplet in (\ref{bulkF}), light modes allowed by the orbifold
BC are written.  Among them, $Q_{\alpha L}$ ($\alpha=1,2,3$) are made 
heavy by coupling with $\hat \chi_{\alpha R}$ on the Planck brane.  
The most general $SU(2)_L \times U(1)_Y$ invariant brane mass term is 
given by 
\beqn
&&\hskip -1cm
- i \delta(y) \bigg\{ \sum_{\alpha =1}^3  \mu_\alpha 
\big( \hat \chi_{\alpha R}^\dagger Q_{\alpha L}
- Q_{\alpha L}^\dagger  \hat \chi_{\alpha R} \big) \cr
&&\hskip 1.5cm
+ \tilde \mu \big( \hat \chi_{2 R}^\dagger \, q_L
- q_L^\dagger  \,  \hat \chi_{2 R} \big) \bigg\}  ~.
\label{braneMass}
\eeqn

We demand only that the scale of the brane masses is much
larger than the KK scale.
With this modest ansatz the equations of motion are solved in the  
background of non-vanishing $\theta_H$.  It turns out that the low energy
spectrum is determined in terms of $c_1$,  $c_2$,  $\theta_H$,
and the ratio $\tilde \mu/ \mu_2$.    The value of each $\mu_\alpha$ 
or $\tilde \mu$ is irrelevant, provided that 
$\mu_\alpha^2, \tilde \mu^2 \gg m_\KK$.
With these brane mass interactions the lightest modes in 
 $Q_{\alpha L}$ ($\alpha=1,2,3$) acquire masses of $O(m_\KK)$.

The top quark mass is generated mainly by the Hosotani mechanism
in the first {\bf 5} multiplet in (\ref{bulkF}).  The bottom quark mass 
is generated by the combination of the Hosotani mechanism 
in the second {\bf 5} multiplet
and  the brane mass terms involving $\mu_2$ and $\tilde \mu$
in (\ref{braneMass}).   For $c_1=c_2 \equiv c < \onehalf$ one finds
\beqn
&&\hskip -1cm 
m_t \sim
\frac{m_\KK}{\sqrt{2}\pi} \, \sqrt{1 - 4c^2} ~  |\sin\theta_H| ~, \cr
\noalign{\kern 5pt}
&&\hskip -1cm
m_b \sim \Big| \frac{\tilde \mu}{\mu_2} \Big| ~ m_t ~.
\label{topbottom}
\eeqn
Given the values of $\theta_H$ and $z_L$, $m_\KK$ is fixed from $m_W$.
Hence $c$ is determined from $m_t$ to be $\sim 0.43$ for 
$\theta_H = \onehalf \pi$ and  $z_L = 10^{15}$.

For the first and second generations similar construction can be done.
For fermions with masses smaller than $m_W$, $c$ becomes larger than 
$\onehalf$.  Indeed, $c \sim 0.65$ and 0.85 for $c$ and $u$ quarks,
respectively.

\subsection{EW symmetry breaking}

The effective potential $V_\eff (\theta_H)$ can be evaluated from the
spectra of the fields in the background $\theta_H$.   
Its evaluation in the RS warped spacetime  was first done
by Oda and Weiler\cite{Oda1}.     A powerful method of
evaluating $V_\eff $ has been developed  by Falkowski.\cite{Falkowski1} 
Concrete evaluation in the gauge-Higgs unification models of electroweak interactions 
in the RS spacetime has been given in refs.\ \cite{Wagner1, HOOS, Hatanaka1}.

\begin{figure}
\includegraphics[width=7.5cm]{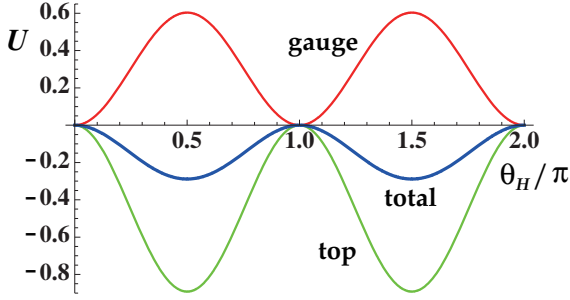}
\caption{Effective potential in the HOOS model.  
$U(\theta_H/\pi) = (4\pi)^2 (kz_L^{-1})^{-4} V_\eff (\theta_H) $ is plotted
at $z_L = 10^{15}$.}
\end{figure}

The effective potential in the HOOS model is depicted in Figure 1.  
In the pure gauge theory without fermions (see the curve denoted 
as ``gauge''),   $V_\eff (\theta_H)$ is minimized at $\theta_H =0, \pi$ 
so that the electroweak symmetry is unbroken.  The contribution from
the top-bottom multiplets with $c \sim 0.43$ denoted as ``top'' in the
figure dominates over the contribution from the gauge fields.  
Contributions from other quarks and leptons are numerically negligible.   
The total effective potential, the curve denoted as ``total'', has the
global minima at $\theta_H = \pm \onehalf \pi$, where the EW symmetry
is spontaneously broken.

The top quark triggers the EW symmetry breaking by the Hosotani 
mechanism in the RS warped space.  One might wonder what would happen 
in flat space.  The effective potential depends on the warp factor $z_L$. 
As $z_L$ decreases, $c$ decreases.  At $z_L \sim 9.4 \times 10^3$, 
$c$ becomes 0 to reproduce $m_t$.   One can further decrease $z_L$
with $c=0$ kept fixed. The relative weight of the top contribution to $V_\eff$
decreases, and at $z_L \sim 900$  the global minima of $V_\eff$ shift
to $\theta_H =0, \pi$ so that the EW symmetry is unbroken.
In other words the EW symmetry is unbroken  in flat space
with the matter content in the HOOS model.

We also note that if all fermions in the bulk belong to the fundamental 
representation of $SO(5)$, then there would be no EW symmetry breaking.
Their contribution to $V_\eff (\theta_H)$ is periodic in $\theta_H$ 
with a period $2\pi$, not $\pi$, and has a minimum either at 
$\theta_H =0$ or $\pi$.  Consequently the total $V_\eff (\theta_H)$ 
has the global minimum either at $\theta_H =0$ or $\pi$ where the 
EW symmetry is unbroken.   If fermions appear in various representations
of $SO(5)$, then the global minimum can be located at $\theta_H$ other
than $0, \pm \onehalf \pi$ and $\pi$ as in the MSW model.

\subsection{Higgs mass and couplings}

The Higgs mass $m_H$ is generated at the one loop level.  It is related to the 
curvature of  $V_\eff$ at the minimum;
\beeq
m_H^2 =  \frac{\pi^2 g_W^2 kL}{4 \,  m_\KK^2}  \, 
\frac{d^2 V_\eff}{d \theta_H^2} \bigg|_{\rm min} ~.
\label{Higgs3}
\eneq
In the gauge-Higgs unification scenario $m_H$ is determined, in essence,  
from the top quark mass $m_t$.   The numerical values are tabulated 
in Table 1 with various values of $z_L$.

\begin{table}[b,t]
\begin{tabular}{ccccc} 
\hline
$z_L$ & $k$ (GeV) & $m_\KK$(TeV) & $c$ & $m_H$(GeV)   \\ \hline 
$10^{17}$ & $5.0 \times 10^{19}$  & 1.58 & 0.438 &  53.5  \\  
$10^{15}$ & $4.7 \times 10^{17}$   & 1.48 & 0.429 &  49.9  \\  
$10^{13}$ & $4.4 \times 10^{15}$   & 1.38 & 0.417 &  46.1  \\ \hline
\end{tabular}
\caption{The Higgs mass $m_H$ in the HOOS model.  $\theta_H= \onehalf \pi$
so that the $WWH$ and $ZZH$ couplings vanish.}
\label{mH-table}
\end{table}

It is seen that the Higgs mass is predicted around 50 GeV
for $z_L = 10^{13} \sim 10^{17}$.  
We stress that this is in no conflict with the  LEP2 bound for $m_H$ which 
states that $m_H < 114 \,$GeV is excluded.  
The crucial observation is that the $ZZH$ coupling vanishes at 
$\theta_H = \onehalf\pi$ as shown in (\ref{Higgs1}).
The process $e^+ e^- \go Z \go Z H$ does not take place 
at $\theta_H = \pm \onehalf \pi$ so that the LEP2 bound is not applicable.  
The $ZZHH$ coupling, on the other hand, 
is multiplied by a factor $\cos 2 \theta_H$ to the coupling  in the 
standard model as in (\ref{Higgs2}) so that  $e^+ e^-  \go ZH H$ can proceed.  
Light Higgs particles might have been already produced.
It is of great interest that a similar scenario emerges in a version of
MSSM where the lightest Higgs  boson has a different coupling to $Z$ from 
that of the Higgs boson in the 
standard model~\cite{Kane}--\cite{Tobe}.
We remark that in the gauge-Higgs unification scenario 
the light Higgs particle with vanishing $WWH$ and $ZZH$ couplings follows from
the dynamics in the theory, but not by tuning parameters.

The Yukawa couplings are also expected to be suppressed compared with
the values in the standard model.\cite{HNSS}  The dominant decay modes 
of the Higgs particle  at $\theta_H = \onehalf \pi$ would be two $\gamma$ 
decay through a top loop and $b \bar b $ decay.

\subsection{Summary}
 
At the onset of the LHC experiments we have, for the first time in history,
the opportunity for directly seeing the origin and structure of the EW 
symmetry breaking.  It could well force us to go beyond the standard model.
Exciting  scenarios include  supersymmetry,  the little Higgs theory, 
the Higgsless theory, and the gauge-Higgs unification theory.  

The gauge-Higgs unification scenario predicts large deviation from the standard
model in the Higgs sector.  The $WWH$ and $ZZH$ couplings are substantially
suppressed.  The Yukawa couplings are also expected to be suppressed.
In the HOOS model the Higgs mass is predicted rather low with vanishing
$WWH$ and $ZZH$ couplings.  Further tiny violation of the weak universality
is predicted,\cite{HNSS} though experimental detection is difficult.  
Small deviation in the 
$WWZ$ coupling can be measured in the future ILC experiments.

The gauge-Higgs unification scenario needs elaboration and refinement.  
We need a model with quarks and leptons which reproduces the observed
Kobayashi-Maskawa matrix and is consistent with the electroweak
precision measurements.  
The forthcoming experiments at LHC certainly  give us clues in understanding 
the structure of the symmetry breaking
and the origin of the Higgs particle.

\subsection{Acknowledgments}

This work was supported in part 
by  Scientific Grants from the Ministry of Education and Science, 
Grant No.\ 20244028, Grant No.\ 20025004, and
Grant No.\ 50324744.



\def\jnl#1#2#3#4{{#1}{\bf #2} (#4) #3}

\def\Zphys{{\em Z.\ Phys.} }
\def\jssc{{\em J.\ Solid State Chem.\ }}
\def\jpsJ{{\em J.\ Phys.\ Soc.\ Japan }}
\def\ptps{{\em Prog.\ Theoret.\ Phys.\ Suppl.\ }}
\def\PTP{{\em Prog.\ Theoret.\ Phys.\  }}

\def\JMP{{\em J. Math.\ Phys.} }
\def\NPB{{\em Nucl.\ Phys.} B}
\def\NP{{\em Nucl.\ Phys.} }
\def\PLB{{\em Phys.\ Lett.} B}
\def\PL{{\em Phys.\ Lett.} }
\def\PRL{\em Phys.\ Rev.\ Lett. }
\def\PRB{{\em Phys.\ Rev.} B}
\def\PRD{{\em Phys.\ Rev.} D}
\def\PRe{{\em Phys.\ Rep.} }
\def\AP{{\em Ann.\ Phys.\ (N.Y.)} }
\def\RMP{{\em Rev.\ Mod.\ Phys.} }
\def\ZPC{{\em Z.\ Phys.} C}
\def\SCI{\em Science}
\def\CMP{\em Comm.\ Math.\ Phys. }
\def\MPLA{{\em Mod.\ Phys.\ Lett.} A}
\def\IJMPA{{\em Int.\ J.\ Mod.\ Phys.} A}
\def\IJMPB{{\em Int.\ J.\ Mod.\ Phys.} B}
\def\EPJC{{\em Eur.\ Phys.\ J.} C}
\def\PR{{\em Phys.\ Rev.} }
\def\JHEP{{\em JHEP} }
\def\cmp{{\em Com.\ Math.\ Phys.}}
\def\JPA{{\em J.\  Phys.} A}
\def\JPG{{\em J.\  Phys.} G}
\def\NJP{{\em New.\ J.\  Phys.} }
\def\CQG{\em Class.\ Quant.\ Grav. }
\def\ATMP{{\em Adv.\ Theoret.\ Math.\ Phys.} }
\def\ibid{{\em ibid.} }

\def\reftitle#1{}                

\bibliographystyle{aipproc}   

\bibliography{sample}

\IfFileExists{\jobname.bbl}{}
 {\typeout{}
  \typeout{******************************************}
  \typeout{** Please run "bibtex \jobname" to optain}
  \typeout{** the bibliography and then re-run LaTeX}
  \typeout{** twice to fix the references!}
  \typeout{******************************************}
  \typeout{}
 }

\end{document}